\documentstyle[preprint,aps,prl,epsf]{revtex}
\preprint{{\it Applied Physics Letters in press}}
 \begin{document}
 \title{Unipolar spin diodes and transistors}
\author  {M. E. Flatt\'e}
\address {Department of Physics and Astronomy, University of Iowa, Iowa City,
IA 52242}
\author  {G. Vignale}
\address {Department of Physics and Astronomy, University of Missouri, Columbia
MO 65211}
 \maketitle
 \begin{abstract}
 Unipolar devices constructed from ferromagnetic semiconducting materials
 with variable magnetization direction
 are shown theoretically to behave very similarly to nonmagnetic bipolar
 devices such as
 the $p-n$ diode and the bipolar (junction) transistor. Such devices may be
 applicable for magnetic sensing, nonvolatile memory, and reprogrammable
 logic.
 \end{abstract}
 \vfill\eject

\tightenlines

 Until recently the emerging field of magnetoelectronics has focused on
 magnetic metals for conducting components\cite{prinz98}.
 Multilayer magnetoelectronic devices, such as giant
 magnetoresistive (GMR)\cite{GMR88} and magnetic tunnel junction
 (MTJ)\cite{Johnson93,Moodera95,Fert96} devices, have
 revolutionized magnetic sensor technology and hold promise for
 reprogrammable logic and nonvolatile memory applications. The performance of
 these devices improves as the spin polarization of the constituent
 material approaches
 100\%, and thus there are continuing efforts to find 100\% spin-polarized
 conducting materials.

 Doped magnetic semiconductors are a promising direction towards such
 materials, for the band-width of the occupied carrier states is narrow.  For
 example, for nondegenerate carriers and a spin splitting of 100 meV
 the spin polarization will be 98\% at room
 temperature. To date high-temperature ($T_{\rm Curie}>100K$)
   ferromagnetic semiconductors such as Ga$_{1-x}$Mn$_x$As
 are effectively $p$-doped.
  {\it Semimagnetic} $n$-doped semiconductors like BeMnZnSe, however, have
 already been shown to be almost 100\% polarized (in the case of BeMnZnSe
 in a 2T
 external field at 30K)\cite{BeMnZnSe}. Both resonant tunneling
  diodes (RTDs)\cite{OhnoRTD}
 and light-emitting diodes (LEDs)\cite{OhnoLED} have been demonstrated which
 incorporate one
 layer of ferromagnetic semiconductor. It is inevitable that
  devices incorporating multiple layers of ferromagnetic
 semiconducting material will be constructed.

 Motivated by this possibility we have investigated the transport
 properties of specific device geometries based on
multilayers of spin-polarized unipolar  doped
semiconductors.
 Previous theoretical work in this area includes spin transport in {\it
homogeneous}  semiconductors\cite{FB,AV} and calculations of spin filtering
effects in superlattices\cite{Egues}.
Our interest here is on the nonlinear
transport
 properties, particularly the behavior of the {\it charge} current,
of two and three-layer heterostructures. We
 focus on two device geometries, which for simplicity we will assume are
 uniformly $n$-doped ($p$-doped devices behave similarly, but with
 opposite sign of the charge current). The first, which will be referred to
 as a spin diode, consists of two layers with
 antiparallel majority carrier spin
 polarization and is in many ways similar to the MTJ devices based on metals.
  The second is a three-layer
 configuration with
 alternating majority carrier spin polarization, which will be referred to as
 a spin transistor.  These devices could function in a similar way to GMR
or MTJ devices, in which
 the resistance of the device changes due to a change in the magnetization
 direction of one layer. As the parallel configuration (or ``low-resistance
state'') would be of higher resistance than that of the GMR or MTJ devices,
these
devices would match better the typical impedance levels of conventional
semiconducting technology.

 The  aspects of these devices we will emphasize in this Letter, however,
are the presence of {\it charge} current gain in the spin transistor and
the sensitivity of
this gain to magnetic field.
Specifically, the $I-V$ characteristics --- unlike those of devices based
on magnetic metals
 --- are
inherently nonlinear.  This suggests new modes of operation of these devices
in reprogrammable logic, nonvolatile memory,  and magnetic sensing.

 In our presentation of the current-voltage characteristics of the spin
transistor
 we will frequently allude to a fundamental analogy
 between unipolar ferromagnetic semiconductors and nonmagnetic bipolar
 materials. This analogy is best visualized in the relationship between a
spin diode
 and the traditional $p-n$ diode.
 Shown in Figure 1(a) are the band edges of the conduction and valence band
 for a traditional $p-n$ diode in
 equilibrium. The quasifermi levels are shown as dashed lines.
 To assist in exploring the analogy with the spin diode, Figure 1(b) shows
 the energies of the elementary carriers
 in those bands: conduction electrons and valence holes. This unfamiliar
 diagram is obtained merely by noting that the energy of a hole in the
 valence band is the
 negative of the energy of the valence electron (relative to the chemical
 potential). We introduce Fig.~1(b) in order to point out the similarities
 with the band edges for the spin diode. Shown in Fig.~1(c) are those band
edges,
 which are also the carrier energies. Just as for
 the $p-n$ diode, in the unipolar spin diode the majority carriers on one
 side are the minority carriers on the other side.

 A major difference, however, is that the two types of carriers in the
 $p-n$ diode have opposite charge, whereas in the spin diode they have the
 same charge. One implication of this is that in the $p-n$ diode the
 interface between the layers is a {\it charge} depletion layer whereas in the
 spin diode the interface is a {\it spin} depletion layer.
The strong spin-density gradient is  maintained by the gradient of a
self-consistent exchange field \cite{footnote}.
 Another major difference resulting from the charges of the carriers is the way
the carrier energies shift under bias.

 In the $p-n$ diode under forward bias
 the barriers for both valence hole and conduction electron transport across
 the junction are reduced. As shown in Fig. 1(d,e) this leads to an
 increase in the conduction electron current to the left and the valence
 hole current to the right. Because the carriers have opposite charge, both
 increases result in an increased charge current to the
 right. For the spin diode only the barrier for spin up electrons moving to
 the left is reduced --- the barrier for spin down electrons moving to the
 right is increased. The charge current is thus directed to the right and
 the spin current to the left. Under reverse bias the barriers for carrier
 transport are both increased in the $p-n$ diode (Fig. 1(g,h)), yielding
 rectification of
 the charge current. For the spin diode (Fig. 1(i)), as before one barrier
 is reduced and the other increased. Thus the charge current is not
 rectified but the spin current is.  Applying analogous assumptions to the
 Shockley assumptions for
 an ideal diode (the validity of these assumptions will be discussed
 below), we find the charge current density $J_q$ and the spin current
 density $J_s$ depend
 on the voltage $V$ according to:
 \begin{eqnarray}
 J_q &=& 2qJ_o{\rm sinh}(qV/kT),\\
 J_s &=& 2\hbar J_o{\rm sinh}^2(qV/2kT),
 \end{eqnarray}
 where $J_o = Dn_m/L_m$, $q$ is the electron charge, $V$ is the voltage, $k$
 is Boltzmann's constant, $T$ is the temperature, $\hbar$ is the Planck's
constant, $D$ is the diffusion
 constant, $n_m$ is the minority carrier density, and $L_m$ is the minority
 spin diffusion length. The resulting spin polarization of the current is
 \begin{equation}
 P = (2qJ_s/\hbar J_q) = {\rm tanh}(qV/2kT).
 \end{equation}
 Thus the spin polarization approaches unity as $V$ gets large, and
 approaches 0 for small $V$. The relative directions of the charge and spin
 currents are shown on Figure 1 for the cases of forward and reverse  bias.

 For ease of use as components in integrated circuits, a magnetoelectronic
 device should allow for magnetic manipulation of the charge current gain ---
 to achieve this we describe the spin transistor, shown in Figure 2.
Analyzing this
 structure in a similar way to a bipolar nonmagnetic transistor,
 the collector current density is
 \begin{eqnarray}
 I_C &=& -{qJ_o \over {\rm sinh}(W/L)}[({\rm e}^{-qV_{EB}/kT}-1)  -
 ({\rm e}^{-qV_{CB}/kT}
  - 1)\cosh(W/L)]\nonumber\\
 && - qJ_o[{\rm e}^{qV_{CB}/kT}-1]
 \end{eqnarray}
 and the emitter current is
 \begin{eqnarray}
 I_E &=& -{qJ_o \over \sinh(W/L)}[({\rm e}^{-qV_{EB}/kT}-1)\cosh(W/L) -
 ({\rm e}^{-qV_{CB}/kT}-1)]\nonumber\\
 &&  + qJ_o [{\rm e}^{qV_{EB}/kT}-1].
 \end{eqnarray}
 The base width is $W$, the voltage between emitter and base is $V_{BE}$,
and the
 voltage between collector and base is $V_{CB}$. The base current is
 $I_B = I_E - I_C$. When $W/L$ is small, $I_B \ll I_C$, which is the
 desired situation for transistor operation (current gain $I_C/I_B\gg 1$).
For appropriate values
 of $V_{EB}$ and $V_{CB}$ ($V_{EB}<0$ and $V_{CB}>0$)
 \begin{eqnarray}
 I_C &=& -{qJ_o \over \sinh(W/L)}[({\rm e}^{-qV_{EB}/kT}-1)  + \cosh(W/L)] -
 qJ_o[{\rm e}^{qV_{CB}/kT}-1]\\
 I_E &=& -{qJ_o \over \sinh(W/L)}[({\rm e}^{-qV_{EB}/kT}-1)\cosh(W/L) + 1]  -
 qJ_o.
 \end{eqnarray}

 The ``emitter efficiency" $\gamma$,  defined as the ratio of the majority
spin-direction charge  current $I_{E \downarrow}$ to the total emitter
current $I_E$ \cite{Sze}, is $1 - e^{qV_{EB}/kT}$ and thus very close to one.
However, in  contrast to bipolar nonmagnetic transistors,  the  ``collector
multiplication factor" $M$, defined as the ratio between the full collector
current $I_C$ and the majority spin-direction charge current $I_{C
\downarrow}$\cite{Sze}, is given by
 \begin{equation}
 M = 1+\sinh(W/L){\rm e}^{q[V_{CB}+V_{EB}]/kT},
 \end{equation}
 which is close to $1$ only if $W/L$ is small.

 Thus we have shown that it should be possible to program a logical circuit
which behaves like
 a bipolar logical circuit, using a
 uniformly-doped unipolar magnetic material. The ``$p$''-like regions
correspond to regions
 with the magnetization pointing one way ($\hat z$) and the ``$n$''-like
regions correspond to region with
 the magnetization pointing along $-\hat z$.  Such logical circuits can
include memory circuits, thus
 indicating that nonvolatile memory can be constructed as well. The
orientation of the magnetic domains
 can be straightforwardly performed in a lateral geometry using similar
techniques as for magnetic
 metallic memories\cite{prinz98}. By incorporating both the logical
elements and the connections between them
 in a single architecture, fabrication of such devices should be more
straightforward than a hybrid magnetic
 metal and semiconductor electronic device architecture.

 We now turn to magnetic sensing applications. For GMR and MTJ devices the
sensing is performed by
 allowing the magnetization of one layer to rotate easily in the presence
of an external field,
 and observing the resistance change. Of course the spin diode could
perform this way as well. The
 spin transistor, however, can detect magnetic fields sensitively even when
the magnetization
 direction of the semiconductor layers is unchanged.

 The effect of an external magnetic field on any section of the spin
transistor is principally to shift the minority band edge. If the chemical
potential is pinned by the external circuit the majority band edge does not
move
significantly. Thus the spin transistor is a minority-spin device (in contrast
to  the ``spin field-effect transistor''\cite{Datta}, which is a
majority-spin device).
The collector and emitter currents in the presence of magnetic fields
$B_E$, $B_B$, and $B_C$ applied to the emitter, the base, and the collector
respectively are
 \begin{eqnarray}
\label{IB1}
 I_C &=& -{qJ_0 \over \sinh(W/L)} \left \{({\rm e}^{[-qV_{EB} -g\mu B_B]/kT}-1)
 + \cosh(W/L) \right \}\nonumber\\
 &&- qJ_0\left \{ {\rm e}^{[qV_{CB}+g\mu B_C/kT]}-1 \right \}   \\
 I_E &=&-{qJ_0 \over \sinh(W/L)} \left \{ ({\rm e}^{[-qV_{EB}-g\mu
B_B]/kT}-1)\cosh(W/L) + 1\right \}  - qJ_0.
 \end{eqnarray}
 Here $g$ is the g-factor and $\mu$ is the magnetic moment of the electron.
 These currents depend {\it exponentially} on the  magnetic fields applied to
the base and the collector, but not on the emitter field $B_E$. Materials
such as BeMnZnSe have
 $g$-factors close to 1000\cite{BeMnZnSe}, thus yielding a change in
current of roughly 0.01\% per gauss
 at room temperature, which is still in the {\it linear} region of the
expansion of the exponentials in Eqs.~(9) and (10).  However materials
with still larger $g$-factors may yet be found (typical sensitivity
 of GMR devices is 1\%/gauss). We also
 note that an electrically isolated magnetic field amplifier can be
employed --- namely a small magnetic
 element which is free to rotate in response to the external field, and can
produce a larger field at the spin transistor base layer.

 We now revisit the Shockley assumptions for an ideal diode. These are (1) the
bulk of the voltage
 drop takes place across the depletion region, (2) the Boltzmann
approximation for transport is valid,
 (3) the minority carrier densities are small compared to majority carrier
densities, and (4) no generation
 currents exist in the depletion layer. Assumption (1) causes the greatest
concern.
 The depletion region in the spin diode is very different than that of
 the $p-n$ diode. In the $p-n$ diode the thickness of the charge depletion
region is set
 by the doping levels in the two regions and the band gap of the material.
In
 the spin diode the  spin depletion region is probably a N\'eel wall, and
its thickness is set by
 the ratio between the magnetic anisotropy energy and the magnetic
stiffness. The anisotropy
 energy can be adjusted through shape engineering.

For optimal device performance of spin diodes and transistors the
domain wall should be very thin. In this limit the spin of carriers
 passing through the domain wall will not precess. When the domain wall is
very thick, however, the electron spin will follow adiabatically the
direction of the macroscopic magnetization and the device will behave
like an ordinary metallic conductor, where the voltage drop is uniformly
distributed along the device.  In the general case, there is a finite
probability that the electrons  emerge from the spin depletion region with
their
spins flipped.  We have analyzed this case and find that, due to the high
resistivity of the no-spin-flip channel, the voltage drop still takes place
mostly across the spin depletion region, unless the probability of
no spin flip is utterly negligible.  The latter case can occur if majority spin
orientation carriers from one side of the junction can directly tunnel into the
majority spin orientation band of the opposite side, as opposed to being
thermally excited above the exchange barrier into the minority spin orientation
band.  This process would effectively ``short" the no-spin-flip channel. If
loss
of spin coherence become a serious problem the domain wall can be replaced by a
nonmagnetic region, as is currently done in MTJs. In the presence of the
nonmagnetic region the relevant length is the spin coherence length, which
can be
quite long.

The remaining three assumptions are of less concern.
Assumption (2) commonly holds in semiconductor devices
so long as the applied voltage is not too large. If the spin splitting in
the magnetic regions is sufficiently large compared to the operation
temperature then assumption (3) will hold. Assumption (4) relies on the spin
coherence time greatly exceeding the transit time through the depletion region
(for the spin diode) or the base (for the spin transistor). Measurements of
long spin coherence times in semiconductors near room
temperature\cite{Awsch97,OhnoPRL} indicate
this assumption is reasonable.

 One of us (G.V.) would like to acknowledge the support of the
 National Science Foundation through Grant Nos. DMR-9706788 and DMR-0074959.

\begin{figure}[h]
\centerline{\epsfxsize=5.2in\epsffile[100 50 530 656]{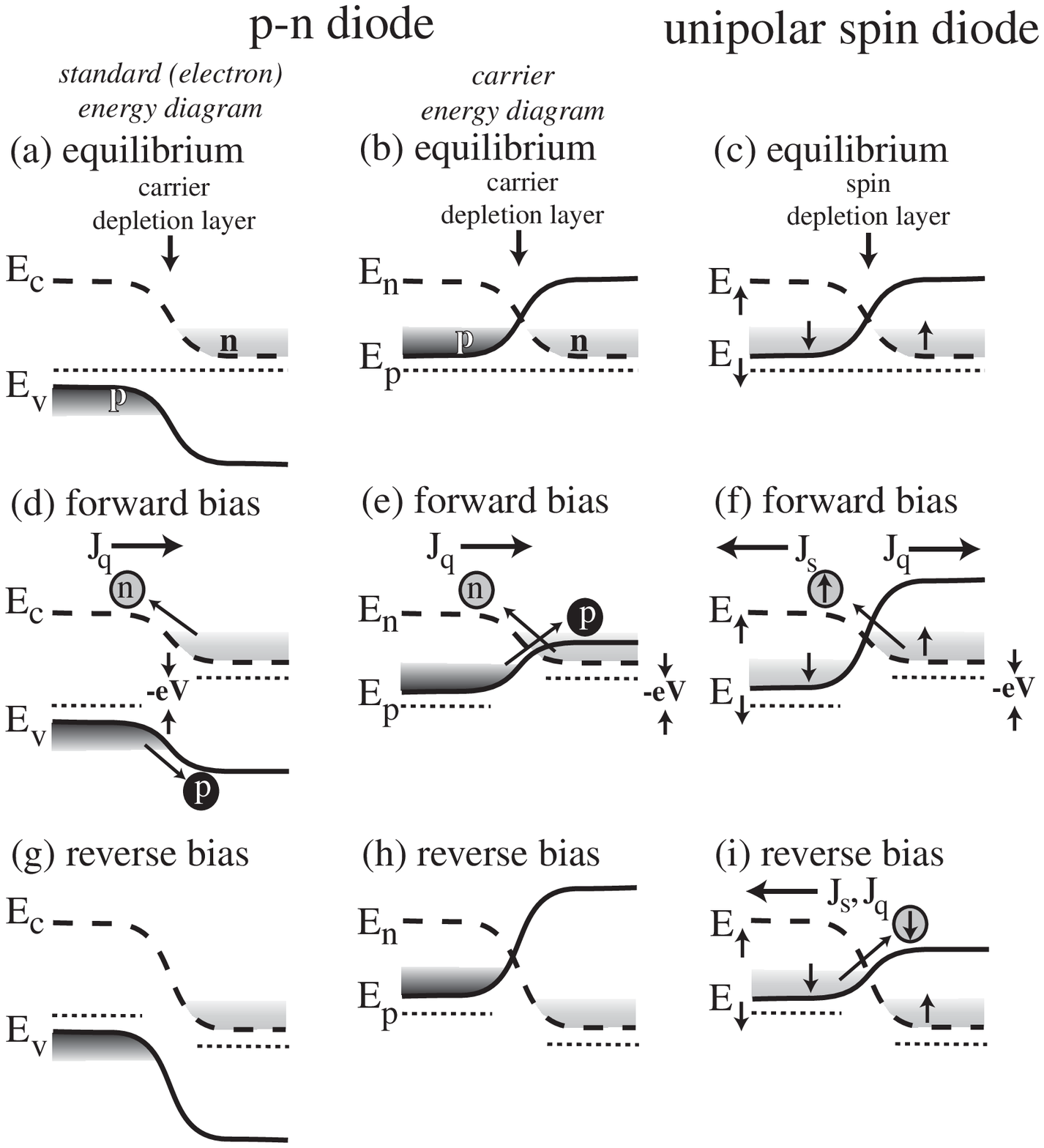}}
\caption[] { Standard and carrier energy diagrams for a traditional p-n 
diode versus unipolar
spin diode under equilibrium conditions (a-c), forward bias (d-f), and 
reverse bias (g-i).}
\label{Fig1}
\end{figure}

\begin{figure}[h]
\centerline{\epsfxsize=5.2in\epsffile[100 350 500 706]{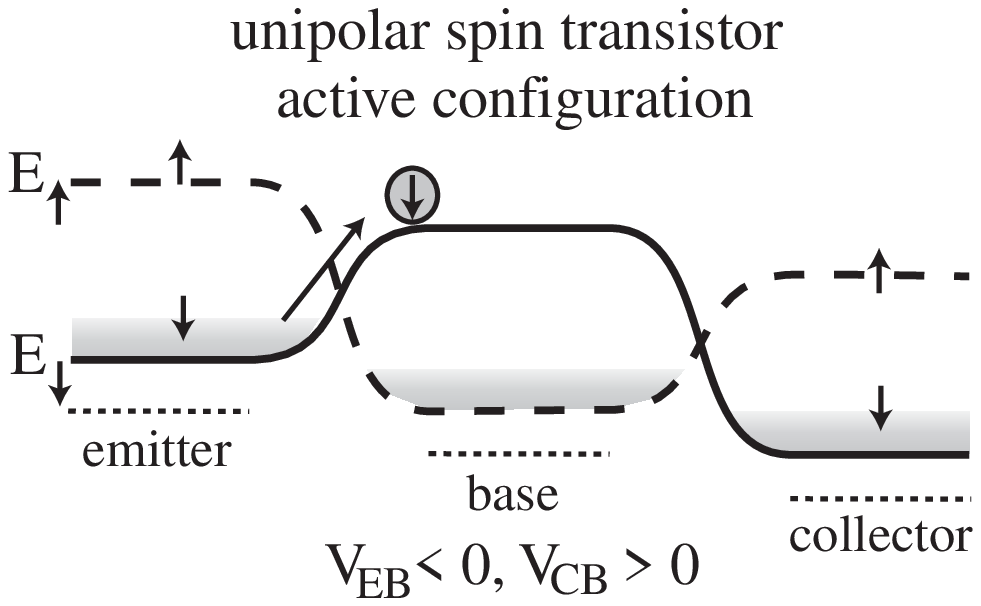}}
\caption[] { Carrier energy diagram for the unipolar spin transistor 
in the normal active configuration. }
\label{Fig2}
\end{figure}

\begin{references}
\bibitem{footnote} The  charge distribution in the spin depletion layer
depends on
the relative size of the screening length (Thomas-Fermi or Debye) and the
width of
the domain wall.  If the former is much larger than the latter the situation
resembles the ordinary depletion layer in a p-n junction:  the charge depleted
from the domain wall is balanced by a screening charge over a screening length
from the domain wall. This  situation  is qualitatively  depicted in Fig.
1(c).
If, on the other hand,  the screening length is much shorter than the width of
the domain wall, then the domain wall region is essentially neutral.
\end{references}
\end{document}